\documentstyle[aps,epsf]{revtex}
\def\openone{\leavevmode\hbox{\small1\kern-3.3pt\normalsize1}}
\newcommand{\la}{\langle}
\newcommand{\ra}{\rangle}
\newcommand{\be}{\begin{equation}}
\newcommand{\ee}{\end{equation}}
\newcommand{\bea}{\begin{eqnarray}}
\newcommand{\eea}{\end{eqnarray}}
\newcommand{\pd}{\partial}
\newcommand{\TEND}[1]{
   {{\raisebox{-.3cm}{$\textstyle\longrightarrow$}} \atop {\scriptstyle{#1}}}}
\newcommand{\APPROX}[1]{
   {{\raisebox{-.3cm}{$\textstyle\sim$}} \atop {\scriptstyle{#1}}}}

\def\gsim{\mathrel{\raise.3ex\hbox{$>$\kern-.75em\lower1ex\hbox{$\sim$}}}}
\def\lsim{\mathrel{\raise.3ex\hbox{$<$\kern-.75em\lower1ex\hbox{$\sim$}}}}
\def\arrl{\mathrel{\raise.3ex\hbox{$\longrightarrow$\kern-1.7em\lower1ex
       \hbox{$\scriptstyle t\rightarrow\infty$}}}}
\def\arra{\mathrel{\raise.3ex\hbox{$\longrightarrow$\kern-1.7em\lower1ex
       \hbox{$\scriptstyle \alpha\rightarrow0$}}}}

\def\lam#1{\lambda_{#1}}
\def\cum#1{\langle\langle {#1}\rangle\rangle}

\def\openone{\leavevmode\hbox{\small1\kern-3.3pt\normalsize1}}
\begin{document} \vspace*{1cm}

\begin{center}
{\Large {\sc  Non-Equilibrium Statistical Mechanics of\\
  Classical and Quantum Systems}}

\vspace{1.3cm}

Dimitri Kusnezov$^1$\footnote{E--mail:
  {\tt dimitri@mirage.physics.yale.edu}},
  Eric Lutz$^1$\footnote{ E--mail:
{\tt eric.lutz@yale.edu}}, Kenichiro Aoki$^2$\footnote{ E--mail:
{\tt ken@phys-h.keio.ac.jp}}  \\
$^1${\small\sl Center for Theoretical Physics, Sloane Physics
  Lab, Yale University, New Haven, CT\ 06520-8120}\\
$^2${\small\sl Dept. of Physics, Keio University,
  4---1---1 Hiyoshi, Kouhoku--ku, Yokohama 223--8521,
  Japan}\\
\end{center}

\vskip 1.8 cm

  {\hspace*{0.3cm} We study the statistical
  mechanics of classical and quantum systems in non-equilibrium
  steady states. Emphasis is placed on systems in strong thermal
  gradients. Various measures and functional forms of observables
  are presented. The quantum problem is set up using random matrix
  techniques, which allows for the construction of the master
  equation. Special solutions are discussed.}

\vspace{2cm}
\section{Introduction}

The efforts to develop statistical mechanics of non-equilibrium
systems are easily traced back to the foundations of the subject.
As statistical mechanics was developed, it was a natural step to
consider how to extend the theory. One can see this in the
handwritten notes of J.W. Gibbs\cite{gibbs}, in which he
formulated the theory of statistical mechanics. Gibbs certainly
expended some effort to understand how he might characterize the
non-equilibrium steady-state. Gibbs certainly puzzled over how to
define entropy for systems in non-equilibrium steady states, as
well as how to define the statistical mechanics of a gas in a
pressure or temperature gradient. However, no solutions were
found. The field has remained active since that time. In the past
ten to fifteen years, the emergence of results in classical chaos
has led to renewed interest and many new ideas have come
about\cite{books}.

Non-equilibrium statistical mechanics is replete with unanswered
questions. While many theoretical techniques have been suggested
to treat transport, a general understanding of the statistical
mechanics is still lacking. In these lectures we survey some
recent results on non-equilibrium steady states in classical and
quantum systems using dynamical boundary conditions.

\section{Some Results from Classical Non-Equilibrium Statistical Mechanics}

\begin{figure}[htbp]
  \begin{center}
    \leavevmode
    \epsfxsize=86mm \epsfbox{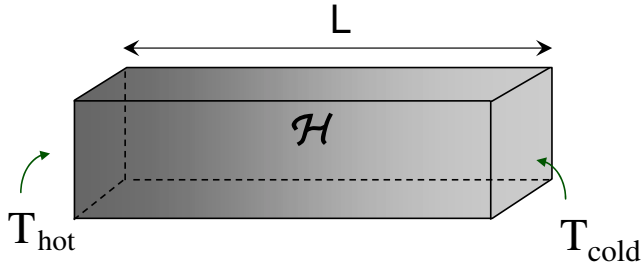}
    \caption{
      Model system of a Hamiltonian $H$ coupled to two thermal
      reservoirs, at temperatures $T_{cold}=T_1^0$ and $T_{hot}=T_2^0$, separated
      by a distance $L$ in the $x-$direction. }
    \label{fig:tgrad}
  \end{center}
\end{figure}

We examine model systems in which a Hamiltonian is coupled to heat
reservoirs at two endpoints, as shown in Fig.~\ref{fig:tgrad}.
The dynamics will be that of the given Hamiltonian, except at the
boundaries where it is coupled to the reservoirs. The boundary
conditions can be implemented in a variety of ways\cite{ak-long}.
Before we discuss quantum systems, we present some results for
classical systems, which summarizes results in Refs.
\cite{ak-long}-\cite{dloss} and references there in.

\vspace{5mm}
 \noindent\underline{Thermal Conductivity.}
\vspace{3mm}

 The transport coefficient in this case is the thermal
conductivity $\kappa$. There are two approaches that are
used to compute $\kappa$. The
first is the Green-Kubo approach, where the non-equilibrium
transport is computed in terms of equilibrium auto-correlation
functions. In this case:
\begin{eqnarray}
  \label{green-kubo}
  \kappa(T)&=&{\frac{1}{T^2}}
  \int_0^\infty\!\!\!\!dt\int\!\!d{x}\,
  \left\langle J({x},t)
  J({x}_0,0)\right\rangle_{\scriptscriptstyle EQ}.
\end{eqnarray}
The heat flow $J$ can be readily computed from the energy-momentum
tensor, $J={\cal T}^{01}$, for heat flow in the $x-$direction.
Alternately, one can compute the conductivity directly in a
non-equilibrium steady state using Fourier's Law. Measuring the
local heat flow $J$ inside the system, and the local temperature
$T(x)$, leads to
\begin{equation}
  \kappa(T) = -\frac{ \langle J
    \rangle_{NE}}{\nabla T(x)}.
\end{equation}
For many systems $\kappa(T)$ behaves as a power law, or can be
approximated by a power law in temperature ranges of interest:
\begin{equation}
  \label{eq:fourier}
  \kappa(T) = \frac{A}{T^\gamma}.
\end{equation}
We should make a few notes concerning Fourier's law.
\begin{itemize}
\item A constant gradient inside a system does not guarantee that Fourier's law
      holds. When $\gamma$ is very small, as in the 1-d Fermi-Pasta-Ulam (FPU)
      $\beta-$model around $T\sim 1$,
      the system can already be too far from equilibrium for Fourier's law to hold,
      yet the nearly constant $\kappa$ still provides a near constant gradient and
      the illusion that the law holds. In the case of the FPU model one finds that
      Fourier's law is not valid even locally when the system has a near constant
      temperature gradient.
\item A strongly-curved temperature profile $T(x)$ (as in Fig. 2) does not mean
      that Fourier's law is not valid. It has been shown that even strongly curved
      profiles can be derived analytically if Fourier's law is satisfied locally.
      So to contrast with the previous point, the curvature of $T(x)$ does in itself
      provide no information on whether Fourier's law is valid.
\item A system does not need to have a bulk limit to satisfy Fourier's law. In the
      1-d FPU model, there is no bulk limit, and $\kappa$ depends on the system size
      $L$. Never the less, Fourier's law is valid even for reasonably strong thermal
      gradients.
\item Fourier's law (2) and the Green-Kubo result (1) have been seen to agree and
      describe the properties of systems even very far from equilibrium, for cases
      where $T(x)$ is strongly curved. Hence linear response theory
      predictions are quite robust and can be expected to hold in strong non-equilibrium
      environments.
\end{itemize}

\vspace{5mm}
\noindent\underline{Temperature Profile.}
\vspace{3mm}

\begin{figure}[htbp]
  \begin{center}
    \leavevmode
    \epsfxsize=86mm \epsfbox{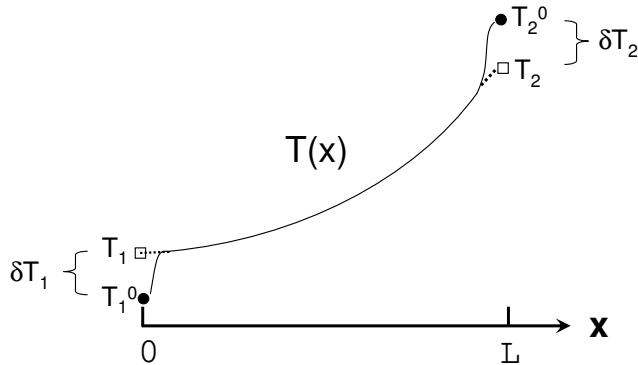}
    \caption{
     Temperature profile of a system such as that depicted in Fig. 1. The boundary
     conditions impose temperatures $T_{cold}=T_1^0$ and $T_{hot}=T_2^0$ on the
edges, $x=0,L$. Just inside the system, there is a boundary jump $\delta T$.
Extrapolating the smooth temperature profile $T(x)$ to the edges defines the temperatures
$T_1$ and $T_2$.}
    \label{fig:tprof}
  \end{center}
\end{figure}

 When we discuss the temperature profile inside,
$T(x)$, we must distinguish the temperatures of the heat
reservoirs, denoted $T_1^0$ and $T_2^0$, and the temperatures just
inside the system $T_1$ and $T_2$.  This is indicated in
Fig.~\ref{fig:tprof}. The boundary jumps $\delta T$ can be
quantitatively understood, as we mention below. The temperature
inside $T(x)$ is given most directly in terms of the extrapolated
temperatures $T_{1,2}$.
We can understand the curvature in the temperature profile using
Fourier's law. When the thermal conductivity can be described by a
power law in the temperature range of interest $(T_1,T_2)$, such
as $\kappa(T)=A T^{-\gamma}$, we can integrate Fourier's law and
re-express the result in terms of extrapolated temperatures
$(T_1,T_2)$:
\begin{equation} \label{t-profile}
  T(x) = \left\{\begin{array}{ll}
    T_1\left[1-\left(1-\left(\frac{T_2}{T_1} \right)^{1-\gamma}
      \right)
      {\frac{x}{L}}\right]^{{\frac{1}{1-\gamma}}},\qquad &
    \gamma\neq1\\
    T_1 \left( \frac{T_2}{T_1}\right)^{x/L}, &\gamma=1\quad.
\end{array}\right.
\end{equation}
Note that $\kappa(T)$ need not have a bulk behavior, nor does it
need to behave globally as a power law to describe $T(x)$ in non-equilibrium
systems.  These formulas are quite robust.

\vspace{5mm}
\noindent\underline{Boundary Jumps.}
\vspace{3mm}

An interesting and often overlooked property of non-equilibrium
steady states is the presence of boundary jumps. In fluids which
are sheared by moving walls, one observes that the velocity just
inside the fluid can be different from the velocity of the wall.
For systems in a thermal gradient, there is an analogous jump in
the value of the temperature just inside the system. This jump,
$\delta T$, is illustrated in Fig. 2.

The jump arises due to a mismatch of the mean free path with the
edge of the system.  It has the general form

\begin{equation}\label{eq:b-jump}
  \delta T_i = T_i^0-T_i=\eta \left.\frac{\partial T}
   {\partial n}\right|_{boundary},
\end{equation}
where $\eta$ is a parameter on the order of the mean free path
$\ell$ and $n$ denotes the normal to the boundary \cite{pk}. This
relation has been verified in two independent manners.

In the first, the boundary jumps $\delta T$ have been  shown to be
linearly related to the normal derivative of the temperature
profile $T(x)$, extrapolated to the edges of the system. The
coefficient $\eta$ can then be extracted. This relation has been
verified in several systems. In the $\phi^4$ model in 1-d, for
example,
\begin{equation}
\eta(T)= (6.1\pm0.5) T^{-1.5\pm0.1}.
\end{equation}
To check if we have the correct understanding, we can make an
estimate of $\eta$ from kinetic theory. Recall that the thermal
conductivity is related to the mean free path by $\kappa\simeq C_V
c \ell$ by a standard kinetic theory argument. For systems where
$C_V$ and $c$ are largely temperature independent, such as in
various lattice models, we expect $\kappa\sim \ell$.
Dimensionally, the coefficient $\eta$ should be on order of the
mean free path $\ell$, so that $\eta\sim\kappa$. Both the
magnitude and power law behavior are consistent with the thermal
conductivity obtained in the $\phi^4$ model in 1-3 D as well as
the FPU model.

An independent verification of the behavior of the jumps can be
made by studying how the jumps depend on the heat flux. We let
$\eta=\alpha\kappa$, where $\alpha$ is a constant to be
determined.  From Eq. (\ref{eq:b-jump}), we can then associate the
heat flux with the right side of Eq. (\ref{eq:b-jump}):
\begin{equation}\label{t-jumps}
  T_i-T_i^0 \simeq \alpha \langle J \rangle_{NE}
          \sim    \alpha (T_2^0-T_1^0)\frac{\kappa(T_{av})}{L}
          + \cdots.
\end{equation}
$\alpha$ is a coefficient which measures the efficacy of the
boundary conditions. The linear relation between $\delta T$ and
$\langle J \rangle_{NE}$ has been verified. $\alpha$ has been
extracted for a variety of boundary thermostats, and can show
strong dependence on those boundary conditions. The understanding
of these jumps together with the temperature
profile~(\ref{t-profile}) provides a complete description of
$T(x)$ in terms of the boundary temperatures, $(T_1^0,T_2^0)$.

\vspace{5mm} \noindent\underline{Observables.} \vspace{3mm}

If we probe an observable somewhere in the system in the
non-equilibriums steady state, can we understand generically how
we expect these to behave?  It is a natural idea that a physical
observable $\cal A$ will deviate from its value in local
equilibrium as we move further away from equilibrium.  In our
case, a temperature gradient $\nabla T$ provides a measure of how
far we are from equilibrium. The deviation of an observable
$\delta {\cal A}$ from its local equilibrium value is expected to
behave as
\begin{equation}
  \label{o-le}
  \delta_{\cal A}\equiv
  {\delta {\cal A}\over{\cal A}}=
  C_{\cal A}\left(\nabla T\over T\right)^2
  +C'_{\cal A}\left(\nabla T\over T\right)^4
  +\ldots
\end{equation}
This expansion is in even powers since the result should be
insensitive to which side of the box is taken as the hot side and
which is cold. (Using Fourier's law, we can equally well expand in
terms of the current $J$.) While this seems natural, there has
been some contention on analogous expansions in sheared fluids.
There non-analytic dependencies on the shear rate have been
observed in some numerical simulations, but the situation has not
been entirely clarified\cite{shears}. The expansion coefficients
$C_{\cal A},C'_{\cal A},\ldots$ are in principle dependent on
physical quantities, such as $T$ and $L$, the size of the lattice
in the direction of the gradient. If the departures from local
equilibrium are local, then the the coefficients would be expected
to be independent of $L$. It has been shown that even in systems
with bulk transport behavior, the coefficients do indeed depend on
the system size.

\vspace{5mm} \noindent\underline{Far from Equilibrium Spatial
Dependence.} \vspace{3mm}

The expansion in (8) has the advantage that we can combine it with
(4) and obtain the spatial information on how observables vary
from local equilibrium values within a system:

\begin{equation}
  \label{o-le1}
  {\delta {\cal A}\over{\cal A}}=
  C_{\cal A}\left(J\over \kappa T\right)^2
  =C_{\cal A}\left(\frac{1}{a+bx}\right)^2
\end{equation}
where
\begin{equation}
 a=\frac{T_1^{1-\gamma}A}{\langle J\rangle_{NE}},\qquad
 b=\gamma-1 .
 \end{equation}
So knowing $C_{\cal A}$, we can also predict the spatial variation
of the non-equilibrium observable ${\cal A}$. This has been tested
and verified in classical lattice models in $d=1-3$ dimensions.

The form of (9) brings to light an interesting relation to coarse
graining. Let us assume we have an observable which is locally
non-equilibrium, so that $\delta A/A$ at some point $x'$ is
non-zero. Using traditional arguments, we would take a larger box
containing $x'$ and assume that on the larger scale, the behavior
is closer to local equilibrium. In examining (9) we see that it is
a positive definite function of position (up to the overall sign
of $C_{\cal A}$). So any averaging over a larger box will not
necessarily converge to $\delta A/A=0$, required for local
equilibrium.

\vspace{5mm} \noindent\underline{Breaking of Local Equilibrium.}
\vspace{3mm}

{\it Local equilibrium} is a property which is usually invoked to
allow application of equilibrium physics to a problem of interest.
It is seldom quantified. Using the expansion (8), it is possible
to quantify the `quality' of the local equilibrium assumption. The
local temperature at position ${\bf x}$ in the model of Fig. 1 is
given by $T_k=\langle p_k^2\rangle$. A natural measure for the
deviations from local equilibrium is the deviation of the momentum
distribution from the Maxwellian distribution. Consider

\begin{equation}
 {\cal A}=\langle p^4\rangle.
 \end{equation}
 In local equilibrium, we expect a Gaussian distribution of
 momenta, thermalized at the local temperature $T(x)$ (or $T_k$), so that

\begin{equation}
 {\cal A}_{eq}=\langle p^4\rangle_{eq}=3 \langle p^2\rangle^2 =3 T(x)^2.
 \end{equation}
Hence a measure of local equilibrium is given by

\begin{equation}
 \delta \langle p_k^4\rangle=\langle p_k^4\rangle-3\langle p_k^2\rangle^2
 =\langle\langle p_k^4\rangle\rangle,
 \end{equation}
where  $\cum{p_k^4}$ is the cumulant at location $k$. In
principle, we can examine the behavior of other cumulants are
measures of the breaking of local equilibrium, such as
$\cum{p_k^6} = \langle p_k^6\rangle -
  15 \langle p_k^4\rangle   \langle p_k^2\rangle
  + 30 \langle p_k^2\rangle ^3$. It is convenient to normalize the cumulants
  by the local temperature,  $\cum{p^n}/T^{n/2}$. These provide a
quantitative measure on  how far we are from local equilibrium. In
local equilibrium, $\cum{p^n}=0$ ($n>2$).

Here a few properties which have been observed in lattice models:
\begin{itemize}
\item Contrary to naive intuition, the steepest gradient in Fig. 2
does {\it not } lead to the system being furthest from local
equilibrium.  In fact, the converse is true --- the system is
furthest from equilibrium in the flattest region.
\item It has been found that $\nabla T/T$
provides a good measure of how far we are from equilibrium and that
 the cumulants behave as
\begin{equation}
  \label{le-breaking}
  \delta_{\scriptsize LE}=\frac{ \cum{p^4}}{3T^2}=
  C_{LE} \left(\frac{\nabla T}{T}\right)^2.
\end{equation}
In the $\phi^4$ and FPU models,$C_{LE}^\phi =1.1(8)L^{0.9(2)} \
(T=1)$ and $C_{LE}^{FPU}=4.3(4)L^{0.99(2)} \ (T=8.8)$.  There
seems to be a weak $T$ dependence for $C_{LE}$ which is difficult
to establish.
\item  These results are
consistent with $d>1$ in lattice models at the same temperatures.
\item Using $C_{LE}^\phi$, $C_{LE}^{FPU}$, one can predict the shape of
$\cum{p^4}/T^2$ in the system using (9), which has been verified
numerically.
\item As the system moves away from
equilibrium by increasing the difference in the boundary
temperatures, each point in the interior deviates from local
equilibrium in a predictable manner, {\it without} any threshold.
Away from equilibrium, local equilibrium is an approximation that
is quite good for small gradients since the deviations from it
only vary as $(\nabla T)^2$.
\item Similar results hold for  higher momentum cumulants.
\end{itemize}

There are many open questions here that related to measurements in
lattice models. The naive expectation would be that since the
gradients and the cumulants are local, their relationship should
not depend on the system size $L$. There does seem to be size
dependence, even in systems with bulk transport, many mean free
paths away from the boundaries.

\vspace{5mm} \noindent\underline{Corrections to Linear Response.}
\vspace{3mm}

There is an interesting relation between linear response predictions for
transport coefficients and the notion of local equilibrium. For instance,
if there is a departure of a predicted transport coefficient from a measured
value for a system far from equilibrium, is this an indication that {\it (a)}
corrections to linear response are needed or {\it (b)} local equilibrium
is no longer a valid assumption. If the latter is true, temperature can no
longer be  unambiguously defined. Near global equilibrium, Fourier's law
is satisfied {\it globally} so that there is a constant thermal gradient:
\begin{equation}
J_0=-\kappa(T) (T_2-T_1)/L\simeq -\kappa(T) (T_2^0-T_1^0)/L.
\end{equation}
As the temperature difference increases, curvature in $T(x)$
usually develops, and Fourier's law is satisfied locally. For
$\kappa\sim T^{-\gamma}$,  Fourier's law can be integrated to
obtain the next leading order correction to $J_0$, due to
curvature in $T(x)$. Denote this heat flow, $J_{LR}$, which
includes the curvature correction. It behaves as:
\begin{equation}
  \label{lr-breaking}
  J_{LR}= J_0\left[ 1 + \frac{\gamma(\gamma+1)}{24}L^2
  \left(\frac{\nabla T}{T}\right)^2+\cdots\right].
\end{equation}
Notice that the curvature correction to $J_0$ behaves as $L^2$.

Using numerical simulations on lattice models, one can push the system
very far from equilibrium and examine how $J$ behaves. It is found that
as the gradient
increases, the energy that can be pumped through the
system becomes less than that predicted by linear response theory
{\it even when it is applied
  locally}. The `violation' of linear response is defined as $\delta_{LR}$ and
is found to behave as
\begin{equation}
  \label{linear-response}
  \delta_{\scriptsize LR}=  {J-J_{LR} \over J_{LR}} =
  C_{LR}\left(\frac{\nabla T}{T}\right)^2.
\end{equation}
In the $\phi^4$ and FPU models, it is found that
$C_{LR}^\phi=-4(3) L ^{1.0(2)}\ (T=1)$ and
$C_{LR}^{FPU}=-6.6(8)L^{0.9(1)}\ (T=8.8)$.
Notice that this will naively give rise to decreasing
current $J$ with increasing gradient. However, when the gradient is
this large, the higher order terms in $(\nabla T/T)$ becomes as
important.  It is believed that the current will eventually saturate
under extremely large gradients.

A very interesting picture emerges. Within error, the violation of
linear response and local equilibrium are closely connected,
occurring in the same manner:
\begin{equation}
  \delta_{LE}\sim \delta_{LR}.
\end{equation}
Further, local equilibrium and linear response have no threshold, and break
down in a similar manner as the system moves away from global thermal
equilibrium.

\vspace{5mm} \noindent\underline{Non-Equilibrium Equation of
State.} \vspace{3mm}

For a simple one-component theory, the equation of state will have a
simple form such as $P_{eq}(T)$ or $E_{eq}(T)$, where $P$ and $E$
are the equilibrium pressure and energy density. In the non-equilibrium
steady state, one expects new variables to emerge. For thermal gradients,
$\nabla T$ will become an independent variable. What is surprising, is that
even in systems with bulk behavior, there can develop a system size
dependence for the non-equilibrium 'equation of state'. For $\phi^4$ and
FPU lattice models,
it is found that
\begin{equation}
  \label{noneqP}
 P(T,\nabla T,L) = P_{eq}(T)
 \left[ 1+ C_P\left(\frac{\nabla T}{T}\right)^2\right]
\end{equation}
where $C_P^\phi=1.5(1.2)L^{0.9(2)} \ (T=1)$, $C_P^{\scriptstyle
  FPU}=4.1(6)L^{0.30(4)}\ (T=8.8) $. Notice that the $L$ dependence makes
the equation of state {\it non-local} even though it is measured locally,
in a system with bulk behavior. If we study the energy density, we
have a similar result:
\begin{equation}
 E(T,\nabla T,L) = E_{eq}(T)
 \left[ 1+ C_E\left(\frac{\nabla T}{T}\right)^2\right]
\end{equation}
where $C_E^\phi=0.5(3)L^{0.9(2)}
\ (T=1)$, $C_E^{\scriptstyle
  FPU}=1.7(7)L^{0.3(1)}\ (T=8.8)$. In contrast,
Extended Irreversible Thermodynamics (EIT), predicts a local
behavior for $P-P_{eq}$ in contrast to our observations.

\vspace{5mm} \noindent\underline{Dimensional Loss.} \vspace{3mm}

A consistent picture that emerges from classical simulations of
non-equilibrium steady states is that the accessible volume of
phase space contracts onto a fractal set. The dimensionality is
obtained from the Kaplan-Yorke estimate, which involves
calculating the full spectrum of Lyapunov exponents.  There is a
corresponding loss of dimension $\Delta D$. An important question
that arises in non-equilibrium systems is whether the dimensional
loss is extensive, and persists in the bulk, or simply a
manifestation of low dimensionality. In systems {\it with bulk
behavior}, we expect $\Delta D$ to behave ``extensively''. That
is, $\Delta D$ should remain relatively the same under the same
local non-equilibrium conditions when we change the size of the
system. Extensivity has been shown in the following form:
\begin{equation}
  \label{dj}
  {\Delta  D\over V_{in}} = C_D J^2
\end{equation}
where $V_{in}$ is the volume of the system inside, excluding the
thermostatted degrees of freedom, while $D$ includes all degrees
of freedom. The coefficient is given by
\begin{equation}
  \label{dj1}
  C_D=    {1\over\kappa\lambda_{max}^{eq}T^2}
  \left(1+{2\alpha\kappa\over V_{in}}\right)
  {{V_{in}\rightarrow\infty}\atop \longrightarrow}
  {1\over\kappa\lambda_{max}^{eq}T^2}
\end{equation}
Here $\lambda_{max}^{eq}$ is the maximum Lyapunov exponent of the
full system plus bath, in thermal equilibrium. Some results are
\begin{itemize}
\item Extensivity is explicitly related to the
macroscopic transport properties through entropy production, which
leads to (22). The first systematic study of the role of dynamical
thermostats on dimensional loss was also shown in [8].
\item While it was emphasized that the
extensivity of $\Delta D$ is not compatible with local equilibrium
so that it is questionable\cite{cels}. As was mentioned above,
violations of local equilibrium appear in the same manner as
dimensional loss. Hence there is no conflict.
\item  An estimate of dimensional loss in an ideal gas gives an idea
of the magnitude of the effect. Using  the standard estimates of
$\lam{max}$\cite{books} yields $C_D\simeq[9\rho
v^2\ln(4l/d)/2]^{-1}$, where $\rho$ is the density, $v$ is the
average particle velocity, $l$ is the mean free path and $d$ is
the particle diameter. For $\nabla T/T\sim0.01\,$m$^{-1}$, $\Delta
D\sim10^9$\,${\rm m}^{-3}$ at room temperature --- quite large,
yet far smaller than the total number of degrees of freedom.

\item The results are satisfying from the physics point of view. $\Delta
D$ pertains to the whole system, it includes the temperature
profile which is curved in general, boundary temperature jumps and
the various types of thermostats. Yet, $\Delta D $ can be related
to macroscopic transport with the thermostat dependence cleanly
separated into $\lam{max}$. Furthermore, $\Delta D$ has extensive
behavior with respect to the internal volume wherein the the
system is manifestly in non-equilibrium. We have seen that $\Delta
D$ and $\lam{max}^{eq}$ both depend strongly on the thermostat
used, while their product $\Delta D\lam{max}^{eq}$ is thermostat
{\sl independent} and related to macroscopic quantities.
\item $\lam{max}^{eq}$ is {\sl not} unique: In global
thermal equilibrium, different choices of heat-baths can lead to
very different values. The result is that dimensional loss is not
unique either, only the product behaves macroscopically and can be
related to thermodynamic quantities
\end{itemize}

\vspace{5mm} \noindent\underline{Non-Equilibrium Temperature Renormalization.}
\vspace{3mm}

In expansions such as (8), it is natural to consider whether other
powers are important, such as  $(\nabla^n T)/T$. In the region
where the temperature dependence of the thermal conductivity  can
be described by a power law, $\kappa= A T^{-\gamma}$, one can show,
using Fourier's law, that

\begin{equation}
\frac{\nabla^n T}{T} = \frac{(\nabla T)^n}{T^n}
(n-1)!\prod_{k=1}^{n-1} \left( \gamma - \frac{k-1}{k}\right).
\end{equation}
While Fourier's law holds only close to local equilibrium, the
deviations from it is of order $(\nabla T/T)^2$ so that the
difference is a higher order correction in the expansion in
$(\nabla T/T)$. Hence the expansion in (8) is sufficiently
general.

When local equilibrium is broken,  the definition of $T$ is no
longer unique. For the observables discussed above, it is possible
see that certain redefinitions of $T$ leave the observables
invariant to leading order. For the generic redefinition
\begin{equation}
 T=T'+ \nu\left(\frac{\nabla T}{T}\right)^2,\qquad  C'_{\cal A}=C_{\cal
  A}+\nu\left(\frac{dA}{dT}\right),
  \end{equation}
the non-equilibrium deviations of $J$, $P$, $E$ are affected
covariantly. Further, the local equilibrium violations seen in
$\cum{p^n}/T^{n/2}$ are {\it
  invariant} under such redefinitions, up to the order we
consider.  The physics, of course, is invariant under any
redefinition in temperature.

\vspace{5mm} \noindent\underline{Summary of Classical Results.}
\vspace{3mm}

We have seen that we can begin to understand the behavior of
systems both near and far from equilibrium.  Certainly many
questions remain, including the behavior of phase transitions in
strongly non-equilibrium steady-state environments, which seem to
display interesting differences with the equilibrium understanding
of phase transitions[7] . But we would now like to turn to the
treatment of the quantum analogs of the above results.

\section{Dynamical Quantum Heat Baths}

Before we explore the quantum analog of Fig. 1 and the results for
classical non-equilibrium steady states, let us first consider how
to formulate a Hamiltonian description of a heat bath. We would
like to use chaos in a similar manner; the (chaotic) interaction
with the baths should cause the system of interest to thermalize.
We decompose the system+heat bath problem as:
\begin{eqnarray}
H(XP,xp) &=&H_S(XP)+H_1(X;xp) \\
&=&H_S(XP)+\overbrace{H_B(xp)+W(X;xp)} .  \nonumber
\end{eqnarray}
The system, $H_S(XP)$, and the bath, $H_B(xp)$, Hamiltonians are
coupled through the interaction $W(X;xp)$. In this formulation,
for simplicity, the interaction does not include any system
momentum dependence. The density of states of the bath+interaction
is defined by $H_1$ through
\begin{equation}
\rho (E)=\overline{{\rm Tr}\delta (H_1-E)}.
\end{equation}
In order to use chaotic methods, we should make a few remarks
concerning Random Matrix Theory (RMT), which has arisen as the
quantum counterpart to classical chaos.

\vspace{3mm}
 \noindent {\sl What is a
random matrix? } A matrix with random matrix elements
\cite{meh91,guh98}, often taken as Gaussian random numbers. This
is convenience rather than necessity, as the Gaussian model is far
more tractable and well developed than other choices. The
motivation is simple. The Schr\"odinger equation,  $H |\psi_i\ra =
E_i |\psi_i\ra$, cannot be solved for complex many--body systems.
This lead Wigner (1951) to adopt a stochastic description instead
of a dynamical description and to replace the Hamilton operator by
a random operator.

\vspace{.3cm} \noindent {\sl How are the matrix elements of a
random matrix
  $H_{ij}$ distributed?} Assume that
\begin{itemize}
\item $H$ is invariant under ``rotation'' (no preferred basis in Hilbert space),
\item the matrix elements $H_{ij}$ ($1\!\leq\! j\!\leq \!N$) are independent.
\end{itemize}
It can then be shown that  $H_{ij}$ are Gaussian random variables.
Depending on the symmetries of the problem, there are only three
possible ``rotations'' in Hilbert space: orthogonal, unitary and
symplectic transformations. Accordingly, there are  only three
Gaussian random matrix ensembles denoted GOE, GUE and GSE (note
that generalizations exist, for example chiral ensembles in QCD).
For GOE,   $P(\{H_{ij}\}) = K \exp[-\sum_{ij} H_{ij}^2 /\la
H_{ij}H_{ji} \ra]$. In general the variance $\la H_{ij}H_{ji}\ra$
depends on $i,j$.

The classical analogy is the velocity distribution in the kinetic
theory of gases. Newton's equations describing the (large number
of) molecules of an ideal gas cannot be solved. This motivated
Maxwell (1859) to use, for the first time in physics, a
statistical approach to treat a dynamical problem ($\rightarrow$
random vector). He assumed that
\begin{itemize}
\item space is isotropic (no preferred direction in velocity space),
\item the components $v_i$ ($i=1, 2, 3$) of the velocity vector are independent.
\end{itemize}
He then found that the velocities are Gaussian distributed (the
well--known Maxwell distribution), $P(\{v_i\}) = K \exp[-\sum_i
v_i^2/\la v_i^2 \ra]$, with $\la v_i^2\ra =2 kT/M$.

\vspace{5mm} \noindent\underline{Heat Bath Model.}
\vspace{3mm}

We will use the eigenbasis of $H_B(xp)$ to define the matrix elements of
$W$. If we take the interaction to be chaotic, it is reasonable
to choose the statistical ensemble to be defined through $W$ according to the
GOE ensemble. As the Gaussian ensemble is entirely characterized by
its first two moments, this is taken to be in the form:
\begin{equation}
\label{eq2}
\overline{\lbrack W(X)]_{kl}}=0,\qquad \overline{%
[W(X)]_{kl}[W(Y)]_{mn} }=[\delta _{km}\delta _{nl}+\delta
_{kn}\delta _{lm}]{\cal G}_{kl}(X-Y).
\end{equation}
The physics is put into the properties of the correlations. We use
the form:
\begin{eqnarray}
{\cal G}_{ij}(X) &=&\frac{\Gamma ^{\downarrow }}{2\pi \sqrt{\rho
( \varepsilon_i)\rho
(\varepsilon_j)}}e^{-(\varepsilon_i-\varepsilon_j)^2/2
\kappa _0^2}G\left( \frac X{X_0}\right) \\
&\equiv &V_{ij}^2G(X/X_0).\label{eq3}
\end{eqnarray}
This type of parameterization was introduced some time ago in the
study of heavy ion collisions\cite{ko76} and widely used
afterwards \cite{wei80,pey90,bul96,lut99}. The physical motivation
is:
\begin{itemize}
\item The interaction is expected to couple only neighboring energy states
$\rightarrow$ $W=$
random band--matrix with bandwidth $\kappa_0\ll N$. For an ordinary random
matrix (not banded),
$\overline{{{\cal G}_{ab}}^2}$= constant.
\item With increasing energy, the states $|a\ra$ and $|b\ra$  become
more and more complex.
Consequently, their overlap is expected to decrease. This is the
physical origin   of  the
factor containing the density of states of the bath.
\item It is possible that correlations exists between $W$ at different values of
position $X$. This information is included in the correlation function $G(X/X_0)$,
where $X_0$ is the characteristic length scale.
\end{itemize}
We emphasize that the coupling to the random matrix  bath is fully
determined by the variance. The parameters that enter are:
$\Gamma^\downarrow $ --  the spreading width; $X_0 $ --
correlation length measuring how rapidly the system decorrelates
due to the bath; $\kappa_0 $  -- band width of the random matrix
$W$;  $\rho=\rho_0\exp[\beta \varepsilon]$  -- density of states
of the bath   used to define the thermodynamic temperature through
$\beta=1/T=d\log\rho/d\varepsilon$.

\vspace{5mm} \noindent\underline{ Influence Functional Approach.}
\vspace{3mm}

The interaction of the system $H_S$ with a chaotic bath $H_B+W$
can act to thermalize the system.  The evolution equation for the
system can be obtained by integrating over the bath. This is most
readily achieved using the influence functional\cite{if}. The
density matrix of the system+bath evolves according to
\begin{equation}
\rho (t)={\cal J}(t,t^{\prime })\rho (t^{\prime }).
\end{equation}
The evolution operator ${\cal J}$ has a simple path integral
representation which is the squared amplitude:
\begin{eqnarray}
{\cal J}(X,x,X^{\prime },x^{\prime },t|Y,y,Y^{\prime },y^{\prime
},t^{\prime }) &=&K(X,x,t|Y,y,0)K^{*}(X^{\prime },x^{\prime
},t^{\prime }|Y^{\prime
},y^{\prime },0) \\
&=&\int_y^xDx\int_Y^XDX\exp \left[ \frac i\hbar S(X,x)\right] \\
&&\times \int_{y^{\prime }}^{x^{\prime }}Dx^{\prime
}\int_{Y^{\prime }}^{X^{\prime }}DX^{\prime }\exp \left[ -\frac
i\hbar S(X^{\prime },x^{\prime })\right] .  \nonumber
\end{eqnarray}
$S$ is the action for the system+bath. The density matrix can be
expressed in terms of this path integral averaged over the bath
degrees of freedom\cite{if}
\begin{eqnarray}
\rho (X,X^{\prime },t) &=&\int dX_0dY_0\rho _0(X_0,X_0^{\prime
})\int_{X(0)=X_0}^{X(t)=X}{\cal D}X(t)\int_{X^{\prime
}(0)=X_0^{\prime
}}^{X^{\prime }(t)=X^{\prime }}{\cal D}X^{\prime }(t)  \nonumber \\
&&\times \exp \left[ \frac i\hbar \left[ S_0(X(t))-S_0(X^{\prime
}(t))\right] \right] {\cal {L}}(X(t),X^{\prime }(t),t).
\end{eqnarray}
with the influence functional given by:
\begin{equation}
{\cal {L}}(X(t),X^{\prime }(t),t)=\overline{\left\langle i|\left\{
T_a\exp \left[ \frac i\hbar \int_0^tdt^{\prime \prime
}H_1(X^{\prime }(t^{\prime \prime }))\right] \right\} \left\{
T\exp \left[ -\frac i\hbar \int_0^tdt^{\prime }H_1(X(t^{\prime
}))\right] \right\} |i\right\rangle }.
\end{equation}
Here $T$ $(T_a)$ is the (anti) time ordering operator.
At this point we can expand the propagators in the
influence function, and use the statistical properties of the matrix elements.
This allows a systematic approach to include finite temperature
effects into the quantum dynamics. To first order in $\beta =1/T$
one finds

\begin{eqnarray}
{\cal {L}}(X,X{^{\prime }}) &=&\exp \left\{ \frac{\Gamma ^{\downarrow }}%
\hbar \int_0^tds\left[ G\left( \frac{X(s)-X{^{\prime
}}(s)}{X_0}\right)
-1\right] \right\}  \nonumber \\
&&\ \times \exp \left\{ \frac{i\beta \Gamma ^{\downarrow }}{4X_0}\int_0^tds[%
\stackrel{.}{X}(s)+\dot {X{^{\prime }}}(s)]G^{\prime }\left( \frac{X(s)-X{%
^{\prime }}(s)}{X_0}\right) \right\} ,
\end{eqnarray}
where $G^{\prime }(x)=dG(x)/dx$. From this one can derive the
effective time evolution of the density matrix for the
system:

\begin{eqnarray}  \label{eq:evol}
i\hbar \partial _t\rho (X,X{^{\prime }},t) &=&\left\{ \frac{P_X^2}{2M}-\frac{%
P_{X^{\prime }}^2}{2M}+U(X)-U(X{^{\prime }})\right. -\frac{\beta
\Gamma
^{\downarrow }\hbar }{4X_0M}G^{\prime }\left( \frac{X-X^{\prime }}{X_0}%
\right) (P_X-P_{X^{\prime }})  \label{evol} \\
&&\ \ +\left. i\Gamma ^{\downarrow }\left[ G\left( \frac{X-{X^{\prime }}}{X_0%
}\right) -1\right] \right\} \rho (X,X{^{\prime }},t),  \nonumber
\end{eqnarray}
with an arbitrary initial condition $\rho (X,{X^{\prime }},0)=\rho
_0(X,{ X^{\prime }})$. This equation provides the quantum dynamics
of the system at finite temperature\cite{if}. The initial
conditions can be taken as gaussian, $\psi _0(X)=\exp
[-X^2/4\sigma ^2]/[2\pi \sigma ^2]^{1/4}$, so that the initial
density matrix is

\begin{equation}
\rho _0(X,X^{\prime })=\frac 1{\sqrt{2\pi \sigma
^2}}e^{-(X^2+X^{\prime }{}^2)/4\sigma ^2}=\frac 1{\sqrt{2\pi
\sigma ^2}}e^{-(4r^2+s^2)/8\sigma ^2}, \label{gauss}
\end{equation}
where
\begin{equation}
r =\frac{X+X^{\prime }}2, \quad
s =X-X^{\prime }.
\end{equation}

In order to learn about the transport behavior, it is convenient
to define the characteristic function $d(s,k,t)$, whose moments
give the transport coefficients:

\begin{equation}
\rho (r,s,t)=\int \frac{dk}{2\pi \hbar }\exp \left(
\frac{ikr}\hbar \right) d(s,k,t).  \label{dtran}
\end{equation}
Then the cumulants are extracted as follows:
\begin{eqnarray}
\ln d(s,k,t)|_{s=0} =\sum_{n=1}^\infty \frac 1{n!}\left(
\frac{ik}\hbar \right) ^n\langle \!\langle X^n\rangle \!\rangle,
\qquad \ln d(s,k,t)|_{k=0} =\sum_{n=1}^\infty \frac 1{n!}\left(
\frac s{i\hbar }\right) ^n\langle \!\langle P^n\rangle \!\rangle .
\end{eqnarray}
The first cumulant is just the average, $\langle \!\langle
r\rangle \!\rangle =\langle \!r\rangle \!$, the second is $\langle
\!\langle r^2\rangle \!\rangle =\langle r^2\rangle \!-\langle
\!r\rangle \!^2,$ the third is usually referred to as kurtosis,
and so forth. The cumulants are closely related to transport coefficients.

\section{Free Particle Strongly Coupled to the Reservoir}

The master equation can be solved in many limits. To see the effects of the bath,
we take the potential $U(X)=0$. Then density matrix is expressed as:
\begin{equation}
\rho(r,s)=e^{ikr/\hbar}\Psi(s).
\end{equation}
This leads to the simplified master equation:
\begin{equation}
\left\{ \frac{i\hbar k}{M}\frac{\partial}{\partial s} -
i\gamma\hbar X_0 G'\left(\frac{s}{X_0}\right)
\frac{\partial}{\partial s}
-i\Gamma^{\downarrow}\left[G\left(\frac{s}{X_0}\right)-1\right]\right\}\Psi(s)=0
\end{equation}
which can be solved through direct integration. For a general correlation function
$G$:
\begin{equation}
\rho(r,s)=\exp\left\{ \frac{ikr}{\hbar}
-\frac{M\Gamma^\downarrow}{\hbar k} \int_0^s
ds'\frac{1-G}{1-(\gamma X_0 M/k) G'}\right\}.
\end{equation}
Consider some examples.

\vspace{2mm} \noindent\underline{Exponential Correlator: $G(x)
=\exp(-|x|)$} $\quad$ Defining $a=\gamma X_0 M/k$ and
$b=M\Gamma^\downarrow/\hbar k$

\begin{equation}
\Psi(s)=\left(\frac{1+a}{a+\exp(s)}\right)^{-(1+a)b/a}e^{ sb/a}.
\end{equation}

\vspace{2mm} \noindent\underline{Cosine Correlator: $ G(x)
=\cos(x)$} $\quad$  In this case, the solution is of the form
\begin{equation}
  \Psi(s)=\left(\frac{1+(a+\sqrt{a^2-1}\tan{s/2})}
   {1+(a-\sqrt{a^2-1}\tan{s/2})}\right)^b .
\end{equation}
From these explicit solutions to the strongly coupled master equation,
one can compute the coordinate and momentum cumulants and demonstrate how
the quantum system evolves in time in the presence of this type of bath.
As our interest is in the weak coupling limit where the bath looks closer to
an idealized heat bath, we will not pursue this here.

\section{Random Matrix Theory  master equation:\\ weak coupling plus one bath}
\subsection{Hamiltonian}
We  consider a system $S$ coupled to a heat bath $B$. The corresponding Hamiltonian is given by
\be
\label{eq1}
H = H_S + H_{B}+ W \ ,
\ee
where $H_S = P^2/2M + U(X) $ describes the system [a particle
moving in a potential $U(x)$],
$H_{B}$ describes the  bath and $W=X \otimes V$ is the coupling
between the system and the
bath. The coupling Hamiltonian $W$ is taken
linear in the position $x$  of the system and
$V$ is an operator acting on the bath. We denote
by $\varepsilon_a$ the energy eigenstates
of the bath, $H_B|a\ra = \varepsilon_a |a\ra$, and
assume that the heat bath is complex (chaotic).
It is then justifiable to model $V$ by a random matrix:
For a complex system, the states $|a\ra$
are  expected to  be highly  complex and to display random
features; it is thus reasonable  to
assume that  $V_{ab}$ is a stochastic function of $a$ and $b$.

\subsection{Master equation}
\label{sec2}
Our aim in this section is to provide an alternate derivation of a quantum
master equation which gives the time evolution of the
(ensemble averaged) reduced density operator of the system,
$\rho_S(t)= \mbox{tr}_B \rho(t)$, when the coupling is weak.
Here $ \rho(t)$ is the total density operator for system plus
bath. We shall employ a method well--known in quantum optics
(see for instance \cite{car93}).
Starting point of the derivation is the von Neumann equation
for  $ \rho(t)$ written in the Interaction Picture,
\be
\label{eq4}
\frac{d\widetilde{\rho}(t)}{dt} = -i\,[\widetilde W(t), \widetilde{\rho}(t)] \ ,
\ee
with $\widetilde A(t)= \exp(iH_0 t) A \exp(-iH_0 t)$ and $H_0= H_S+H_B$.
Equation (\ref{eq4}) can be formally integrated to give
\be
\label{eq5}
\widetilde \rho(t) = \widetilde \rho(0) -i \int_0^\infty dt' \,
[\widetilde W(t'), \widetilde{\rho}(t')] \ .
\ee
We now substitute this expression for $\widetilde \rho(t)$ inside
the commutator of  (\ref{eq4}) and obtain
\be
\label{eq6}
\frac{d\widetilde{\rho}(t)}{dt} = -i\,[\widetilde W(t), \widetilde{\rho}(0)] -
\int_0^\infty dt' \, [\widetilde W(t),[\widetilde W(t'), \widetilde{\rho}(t')]] \ .
\ee
This equation is still exact. We next  assume that initially the
system and the bath are not correlated and that the latter is in
thermal equilibrium, i.e., $  \rho(0)= \rho_s(0) \otimes \rho_B(0)$
with $\rho_B(0) = Z_B^{-1} \exp(-\beta H_B)$. We also assume
(Born approximation) that the bath remains in equilibrium at
all times so that we can write $\widetilde \rho(t')= \widetilde \rho_S(t')
\otimes \widetilde \rho_B(0)$. Clearly, this approximation is
true provided  the coupling  between the system and the bath is
weak. We now take the trace over the bath and  note  that
$\mbox{tr}_B\widetilde \rho(t)= \exp(iH_S t)\mbox{tr}_B \rho(t)
 \exp(-i H_S t) = \widetilde \rho_S(t)$.  After ensemble averaging
we obtain
 \bea \label{eq601} \frac{d\overline{\rho}_S(t)}{dt}=&
-&\int_0^t d\tau\, \overline K(\tau)\Big\{\widetilde
x(t)\widetilde x(t-\tau) \overline \rho_S(t-\tau) -\widetilde
x(t-\tau) \overline
\rho_S(t-\tau) \widetilde x(t) \Big\} \nonumber \\
&+& \int_0^t d\tau \, \overline K(-\tau) \Big\{ \overline
\rho_S(t-\tau)\widetilde x(t-\tau)\widetilde x(t) -\widetilde x(t)
\overline \rho_S(t-\tau) \widetilde x(t-\tau) \Big\} \ ,
\eea
where we have used  $\widetilde W(t) = \widetilde x(t) \otimes
\widetilde V(t)$ and introduced the bath correlation function
$K(t) =K'(t) +i K''(t) =  \mbox{tr}_B [\widetilde{V}(t)\widetilde{V}(0)
\rho_B(0) ]$.
 Its ensemble average is given by  the Fourier transform of
$\overline{{V_{ab}}^2}$ with respect to $\varepsilon_b$ \cite{lut01},
\be
\label{eq7}
\overline{K}(t) = \int_{-\infty}^{+\infty} d\varepsilon_b\,
\rho(\varepsilon_b)\, \overline{{V_{ab}}^2}\,
e^{i(\varepsilon_a-\varepsilon_b)t} \ .
\ee
Transforming Eq.~(\ref{eq6}) back into the Schr\"odinger Picture yields,
\bea
\label{eq8}
\frac{d\overline \rho_S(t)}{dt} =&-& i\,[H_S,\overline
\rho_S(t)] - \int_0^\infty d\tau \, [x,[\widetilde x(-\tau),\overline \rho_S(t)]]\,
\overline  K'(\tau) \nonumber \\
&- & i\, \int_0^\infty d\tau  \,[x,\{\widetilde x(-\tau),\overline \rho_S(t)\}] \, \overline K''(\tau) \ .
\eea
Equation (\ref{eq8}) is the ensemble averaged master equation we were looking for. Let us
 consider the limit of large bandwidth and high temperature, $1\ll \kappa_0\ll T$.
 Using Eq.~(\ref{eq2}) we find
\be \label{eq9} \overline K(\tau) = \frac{\Gamma}{\sqrt{2\pi}}
\kappa_0 \exp\Big[
-\frac{\kappa_0^2}{2}\Big(\tau+i\frac{\beta}{2}\Big)^2\Big]
\TEND{\kappa_0 \rightarrow \infty} \Gamma\,
\delta(\tau+i\frac{\beta}{2})\TEND{\beta \rightarrow 0} \Gamma \,
\delta(\tau) + i\Gamma \frac{\beta}{2} \, \delta'(\tau) \ , \ee so
that $\overline K'(\tau)= \Gamma \,\delta(\tau)$ and $\overline
K''(\tau) = \Gamma \beta \,\delta'(\tau)/2$. We have accordingly,
\be \label{eq10} \int_0^\infty d\tau \, \widetilde x(-\tau)\,
\overline K'(\tau) = \frac{\Gamma}{2} x  = D x \, \ee where we
have defined the diffusion coefficient $D= \Gamma /2 $, and \be
\label{eq11} \int_0^\infty d\tau \, \widetilde x(-\tau)\,
\overline K''(\tau) = -\Gamma \frac{\beta}{4} \frac{d \widetilde x
(-\tau)}{d\tau}\Big|_{\tau=0} = i\Gamma \frac{\beta}{4}  \,[H_S,x]
= \gamma p \ . \ee Here we have used $[H_S, x] = [p^2, x]/2M= -ip
/M$ and defined the friction coefficient $\gamma = \Gamma\beta
/4M$. Collecting everything, we finally obtain \be \label{eq12}
\frac{d\overline \rho_S(t)}{dt} = -i \,[H_S,\overline \rho_S(t)] -
D [x,[x,\overline \rho_S(t)]] -i \gamma [x,\{p,\overline
\rho_S(t)\}] \ . \ee The master equation (\ref{eq12}) is often
referred to as the Caldeira--Leggett master equation
\cite{cal83,wei99}. It consists of three parts: a von Neumann part
which describes the free motion without the coupling to the
environment, a diffusive part and a dissipative part. Note that
the diffusion and friction coefficients satisfy the Einstein
relation, $D=2M T \gamma$, which expresses the fact that diffusion
of the particle and damping of its energy have a common physical
origin, namely the coupling with the heat bath. The Einstein
relation is an example of the more general
fluctuation--dissipation theorem. We also mention that
Eq.~(\ref{eq12}) is often derived from an oscillator bath model
where the system is coupled to a set of independent harmonic
oscillators (thus an integrable system). It is quite remarkable
that the coupling to a complex random matrix environment leads to
the same equation. This shows that in the limit considered here
(weak coupling, high temperature), the master equation
(\ref{eq12}) is independent of the specific structure of the bath
as well as  of the specific form of the coupling and therefore
universal \cite{lut99}.

\vspace{.3cm}
\noindent
\textbullet { Coordinate representation and semiclassical limit}\\
The coordinate representation of the master equation (\ref{eq12}) reads
\be
\label{eq1201}
\frac{\partial \overline \rho_S(x,x',t)}{\partial t} =
\Big[-\frac{i}{\hbar}\Big(H_S(x)-H_S(x')\Big) - \frac{D}{\hbar^2}\,(x-x')^2 -\frac{i}{\hbar}
\gamma\,(x-x')(p_x-p_{x'})\Big] \overline \rho_S(x,x',t)\ ,
\ee
where for convenience we have reintroduced the constant $\hbar$. This corresponds
to the master equation derived from the influence functional in the limit
\begin{equation}
G(x)\sim 1-\frac{1}{2} x^2.
\end{equation}
The  corresponding classical transport equation is obtained by taking the
Wigner transform of the density matrix
\be
\label{eq1202}
f(q,p,t)= \frac{1}{2\pi\hbar}\int_{-\infty}^\infty dr \exp\Big[-\frac{ipr}{\hbar}\Big]
\rho\Big(q+\frac{r}{2}, q-\frac{r}{2},t\Big)  \ .
\ee
Keeping only terms in leading orders of $\hbar$, this leads to
\be
\label{eq1203}
\frac{\partial f}{\partial t}= -\frac{p}{M}\frac{\partial f}{\partial
  q}+ U'(q)\frac{\partial f}{\partial p}+ 2\gamma
 \frac{\partial }{\partial p}(pf) + D \frac{\partial^2
  f}{\partial p^2}\ .
 \ee
This is the usual Klein--Kramers equation.
\subsection{Relation with the oscillator bath model}
We now turn to a direct comparison of the oscillator bath model and the random matrix
model \cite{lut01}. We still place ourselves in the weak coupling limit. The oscillator
 bath model consists of a particle coupled to a large number of independent harmonic
 oscillators (mass $m_i$ and frequency $\omega_i$). The Hamiltonian of the composite
 system is then (see \cite{wei99} and references therein)
\be
\label{eq13}
H= H_S +\sum_{i=1}^N \left(\frac{p_i^2}{2m_i} + \frac{m_i\omega_i^2}{2}
\left[x_i-\frac{c_i}{m_i\omega_i^2 x}\right]^2\right) \ .
\ee
In this model, the coupling Hamiltonian $W=x \sum_i c_i x_i$ is bilinear
in the position of the system and the the positions of the harmonic oscillators.
Here $c_i$ are coupling constants.  The coupling to the harmonic  bath is fully
characterized by the spectral density function,
\be
\label{eq14}
I(\omega) = \pi \sum_i \frac{c_i^2}{2m_i\omega_i}\,
\delta(\omega-\omega_i)\ .
\ee
In the following we shall relate $I(\omega)$ and the variance
$ \overline{{V_{ab}}^2}$. To this end it is useful to rewrite Eq.~(\ref{eq3}) in the form
\be
\label{eq15}
\overline{{V_{ab}}^2} = \left[ \rho(\varepsilon_a)
    \rho(\varepsilon_b)\right]^{-\frac{1}{2}} \mbox{f}(\omega_{ab}=
  \varepsilon_b-\varepsilon_a)\ ,
\ee where we have introduced the band form factor
$\mbox{f}(\omega)$. It can be shown that with  the form
(\ref{eq15}) the bath correlation function satisfies  the KMS
condition  $ \overline{K}(-t)= \overline{K}(t-i\beta)$ which
defines the thermal equilibrium state of the bath. The bath
correlation function can further be expressed in terms of the
spectral density function as \be \label{eq16} K(t)= \int_0^\infty
\frac{d\omega}{\pi}
I(\omega)\left(\coth\left(\frac{\beta\omega}{2}\right) \cos(\omega
t) - i
  \sin(\omega t) \right)
\ee
We  see  from Eq.~(\ref{eq16}) that  the imaginary part $K''(t)$ of the
correlation function
 is simply the Fourier sine
transform of the spectral density $I(\omega)$. Comparing the imaginary parts
of Eqs.~(\ref{eq8}) and   (\ref{eq16}), we find,
\be
\label{eq17}
I(\omega) = 2\pi \sinh\left(\frac{\beta\omega}{2}\right)
  \mbox{f}(\omega)\ .
\ee This equation provides the desired link between the two
models. It shows that there is a one--to--one correspondence
between the density function $I(\omega)$ of the oscillator bath
model and the variance $\overline{{V_{ab}}^2}$ of the random
band--matrix model. It is important to note that both the
oscillator bath model and the random matrix model, in the limit of
weak coupling we consider here, are Gaussian. This means that in
these two models the dynamics of the bath is is entirely
characterized by the two--point correlation function $K(t)$ and it
is therefore not necessary to consider higher--order correlation
functions.

A  straightforward application of  Eq.~(\ref{eq17}) using the variance  Eq.~(\ref{eq3}) leads to
 \be
\label{eq18}
I(\omega) = \Gamma  \sinh\left(\frac{\beta\omega}{2}\right)
\exp\left[-\frac{\omega^2}{2\kappa_0^2}\right]\ .
\ee
In the limit $\omega \rightarrow 0$, Eq.~(\ref{eq18}) reduces to $I(\omega) \sim M\gamma \omega$
which defines the so--called Ohmic regime. In this regime, as we already saw in the derivation
of the master equation above, the bath correlation function is delta--correlated in
time, $\overline K'(t) = \Gamma \delta (t)$, which  means there are no memory effects.
The process in thus Markovian.  We can  further  easily check that the band form factor
\be
\label{eq181}
\mbox{f}(\omega) = \frac{\Gamma}{2\pi} |w|^{\alpha-1} \exp\left[-\frac{\omega^2}{2\kappa_0^2}\right]\ ,
\ee
yields  $I(\omega) \sim \pi\beta\omega \mbox{f}(\omega)\sim \omega^\alpha$ as $\omega \rightarrow 0$.
This corresponds to fractal (non--Ohmic) environment. Here, contrary to the
Ohmic case, the bath correlation is given by a inverse power law,
\be
\label{eq19}
\overline{K'}(t) =  \frac{\Gamma}{\pi} \Gamma(\alpha) \cos(\frac{\alpha\pi}{2}) \,t^{-\alpha}
\! \ ,\;\;\;\overline{K''}(t) =\frac{\beta}{2}  \frac{d \overline{K'}}{dt} \ ,
\ee
in the limit $1\ll\kappa_0\ll T$.
As a result, there are long--time (memory) effects (the process in now non-Markovian)
and the dynamics of  the particle becomes anomalous. We shall develop that point in
more detail in the next section. But before we would like to show  that the band
form  factor $\mbox{f}(\omega)$ can be directly  related to  the velocity
autocorrelation function (VACF)  $\la v(0) v(t)\ra$ of the system. This is done by  using the first
fluctuation--dissipation theorem
\be
\label{eq20}
C_v[z] = \Big(z+\gamma[z]\Big)^{-1}
\ee
which  relates the Laplace transform $C_v[z]$
of  the
normalized VACF $C_v(t) = \la v(0)v(t)\ra
/ \la v(0)^2\ra$ to  the Laplace transform  of the damping
kernel $\gamma(t)=  \overline K'(t)/M T$ (see below). From Eq.~(\ref{eq7}), we have
\be
\label{eq21}
\gamma[z] = \frac{\beta}{M} \int_{-\infty}^\infty d\omega\,
e^{-\beta \omega}\mbox{f}(\omega) \frac{z}{z^2+\omega^2}\ .
\ee
In the limit $z\rightarrow 0$, the last factor in Eq.~(\ref{eq21}) reduces to a
delta function and we obtain
\be
\label{eq22}
\lim_{z\rightarrow0}\, \gamma[z] = \frac{\pi\beta}{M}\,
\lim_{z\rightarrow0}\, \mbox{f}(z) \ ,
\ee
The final
value theorem of  the theory of  Laplace Transform tells us that
\be
\lim_{t\rightarrow\infty}C_v(t)=   \lim_{z\rightarrow0} z\, C_v[z]
\ee
and we can therefore write
\be
\label{eq23}
\lim_{t\rightarrow\infty}C_v(t) = \lim_{z\rightarrow0} \Big(1+
\frac{\pi\beta}{M}\,z^{-1}\mbox{f}(z) \Big)^{-1}\ .
\ee
We thus see that the long--time behavior of the VACF is determined by the shape of the
band form factor of the origin.

\subsection{Langevin equation}
Consider a system operator $P$ (for instance the momentum operator $p$ of the particle).
Our aim in the present section is
to derive a quantum Langevin equation for this operator in the
limit of weak coupling. The time evolution of the operator $P$ is
given by $P(t) = \exp(iHt) \, P\, \exp(-iHt)$ (Heisenberg
representation). Accordingly,
\be
\label{eq24}
\dot P(t)=  i[H_S(t),P(t)] + i [x(t),P(t)] V(t) \ .
 \ee
This equation is exact. We shall now look for an approximate
expression  for $[x(t),P(t)] = \exp(iHt)\, [x,P]\,\exp(-iHt)$ and
$V(t)=\exp(iHt)\, V\, \exp(-iHt)$. In the limit of weak coupling,
the time evolution operator $\exp(-iHt)$ can be written  in a
Dyson series as,
\be \exp(-iHt) \simeq \exp(-iH_0 t) -i \int_0^t
d\tau \, \exp(-iH_0(t-\tau)) \,W\, \exp(-iH_0 \tau)\ .
 \ee
In lowest order in $V$, this leads to
 \be
\label{eq25}
[x(t),P(t)]= [\widetilde
x(t),\widetilde P(t)] + i \int_0^t d\tau \, [\widetilde
x(t-\tau),[\widetilde x(t),\widetilde P(t)]] \widetilde V(t)\ ,
\ee
and
\be
\label{eq26}
V(t) = \widetilde V(t) + i \int_0^t d\tau \,\widetilde x(t)
\Big(\widetilde V(t-\tau)\widetilde V(t) -\widetilde V(t)
\widetilde V(t-\tau)\Big) \ .
\ee
 Inserting Eqs.~(\ref{eq25}) and (\ref{eq26}) into Eq.~(\ref{eq24})
 we then obtain up to second order in $V$,
\bea
\label{eq206}
\dot P(t)&=& i[H_S(t),P(t)] -\int_0^t d\tau \, [\widetilde x(t-\tau),
[\widetilde x(t), \widetilde P(t)]] \mbox{Re}[\widetilde V(t-\tau)\widetilde V(t)]\nonumber \\
&-& i \int_0^t d\tau\,  \{\widetilde x(t-\tau),[\widetilde x(t),
\widetilde P(t)]\} \mbox{Im}[\widetilde V(t-\tau)\widetilde V(t)]
+i [\widetilde x(t), \widetilde P(t)] \widetilde V(t) \ .
\eea
This equation is valid for any system operator $P$. We now consider the case
where $P=p$ the momentum operator of the particle. Using the commutation
relation,  $[x,p]=i$, we can rewrite Eq.~(\ref{eq26}) in the form
\be
\label{eq27}
\dot p(t) = -U'(x(t)) + 2 \int_0^t d\tau \, \mbox{Im}[\widetilde V(t-\tau)
\widetilde V(t)] \widetilde x(t-\tau) -\widetilde V(t) \ .
\ee
Equation (\ref{eq27})  is almost a Langevin equation. We further  note
that $\widetilde V(t)$ depends on the (thermal) initial conditions of the
 heat bath. This makes the force operator $\xi(t)= - \widetilde V(t)$ a
 fluctuating quantity. Introducing first the thermal average over the bath
 and then  the  average over the random matrix ensemble, we get
\be
 \overline{\la\xi(t) \ra} = 0, \hspace{0.5cm} \overline{\la \xi(t) \xi(0)\ra}=
 \overline{\la \widetilde V(t) \widetilde V(0)\ra}=\overline K(t) \ .
\ee
Moreover, after partial integration, we have
\be
2 \int_0^t d\tau \, \overline K''(-\tau) \overline x(t-\tau) \simeq
-\beta \int_0^t d\tau \,\overline K'(t-\tau) \dot{\overline x}(\tau) \ ,
\ee
where we have used Eq.~(\ref{eq19}).
In the limit of weak coupling and high temperature, the Langevin
equation can therefore be written as
 \be
\label{eq30}
M \ddot{\overline x}(t) +M \int_0^t d\tau \,\gamma(t-\tau)
\dot{\overline x}(\tau) +U'(\overline x(t))= \overline \xi(t) \ ,
\ee
where we have introduced the damping kernel  $\gamma(t)$
which obeys $M T \gamma(t) = \overline{K'}(t) $. This last relation is
often referred to as the second fluctuation--dissipation theorem.
We note  that the Langevin equation is completely determined by the real part
$\overline{K'}(t)$ of the bath correlation function. Introducing the
Riemann--Liouville fractional derivative ($-1\!<\!\lambda\!<\!0$)\cite{sai97},
\be
\label{eq31}
\frac{\pd^\lambda f(t)}{\pd t^\lambda} = \frac{1}{\Gamma(-\lambda)}
 \int_0^t \frac{f(\tau)\,d\tau}{(t-\tau)^{\lambda+1}} \ ,
\ee
 we can  rewrite (\ref{eq30}) in the form of  a fractional Langevin equation \cite{lut01a}
\be
\label{eq32}
M \ddot x(t) + M  \gamma_\alpha  \,\frac{\pd^{\alpha-1}}{\pd t^{\alpha-1}}
\dot x(t) +U'(x(t)) = \xi(t) \ ,
\ee
where we have defined
$\gamma_\alpha =  \Gamma \beta  /(2M  \sin(\alpha\pi/2) )$ and dropped the overline.
The fractional Langevin equation (\ref{eq31}) can be easily solved for the case of a free particle
$U(x) =0$ by making use of the   Laplace transform. We find
\be
\label{eq33}
x(t) = x_0 + v_0   B_v(t) + \int_0^t d\tau \, B_v(t-\tau) \xi(\tau)\ ,
\ee
where $(x_0,v_0)$ are the initial coordinates of the particle and
$B_v(t) = \int_0^t C_v(t') dt'$ is the integral  of the (normalized) velocity
autocorrelation function  $C_v(t)$. Since  $\gamma[z]= \gamma_\alpha z^{\alpha-1}$,
 the Laplace of the VACF is given by
\be
\label{eq34}
C_v[z]=   \frac{1}{z+\gamma_\alpha z^{\alpha-1}} \ ,
\ee
By taking the  inverse Laplace transform,
the velocity autocorrelation function  can be written as
\be
\label{eq35}
 C_v(t) = E_{2-\alpha}(-\gamma_\alpha t^{2-\alpha}) \ .
\ee
Here we have introduced the Mittag--Leffler function $E_\alpha(t) $,
which is  defined by the series expansion \cite{erd55}
\be
\label{eq36}
E_\alpha(t) = \sum_{n=0}^\infty \frac{t^n}{\Gamma(\alpha n+1)} \ .
\ee
The function  $E_\alpha(t) $ reduces to the exponential when $\alpha=1$.
The asymptotic behavior of  the Mittag--Leffler function (\ref{eq36})  for
short and long times is  respectively given by
$\sim \exp(t)$ and $\sim -(t\,\Gamma(1-\alpha))^{-1}$, $0\!<\!\alpha\!<\!1$ and
 $1\!<\!\alpha\!<\!2$. For the velocity
autocorrelation function (\ref{eq35}) this yields a typical stretched
exponential behavior at short times
\be
 \label{eq37}
C_v(t) \sim \exp\frac{-\gamma_\alpha t^{2-\alpha}}{\Gamma(3-\alpha)},
\hspace{1cm}t \ll\frac{1}{(\gamma_\alpha)^{1/\alpha}} \ ,
\ee
and an inverse power--law tail at long  times
\be
\label{eq38}
C_v(t) \sim \frac{t^{\alpha-2}}{\gamma_\alpha \Gamma(\alpha-1)} ,
\hspace{1cm} t \gg\frac{1}{(\gamma_\alpha)^{1/\alpha}} \  .
\ee
  After time integration, we finally get from Eq.~(\ref{eq35})
\be
 B_v(t) = t \,E_{2-\alpha,2}(-\gamma_\alpha t^{2-\alpha}) \ ,
\ee
where  we have used the generalized Mittag--Leffler function
$E_{\alpha,\beta}(t)$  defined as \cite{erd55}
\be
E_{\alpha,\beta}(t) = \sum_n ^\infty \frac{t^n}{  \Gamma( \alpha n+\beta)}\  .
\ee
In the  long--time limit, the generalized Mittag--Leffler function satisfies
$E_{\alpha,\beta}(t) \sim -(t\, \Gamma(\beta-\alpha))^{-1}$.
Accordingly,  $B_v(t)$ exhibits a decay of the form
\be
B_v(t)\sim \frac{t^{\alpha-1}}{\gamma_\alpha \Gamma(\alpha)}\ ,
\hspace{0.7cm}\mbox{ when }  t\rightarrow \infty \ .
\ee
We emphasize  that  the solution (\ref{eq33}) of the fractional Langevin
equation in   the force  free case is completely specified by  the knowledge
of the function  $B_v(t)$  .

The mean displacement and the mean--square  displacement  are
readily deduced from Eq.~(\ref{eq33}). We find
\be
\label{eq39}
\la x \ra = x_0 + v_0 \,  t \,E_{2-\alpha,2}(-\gamma_\alpha t^{2-\alpha})
\APPROX{t\rightarrow \infty}  \frac{v_0}{\gamma_\alpha}
\frac{t^{\alpha-1}}{\Gamma(\alpha)}
\ee
and
\be
\label{eq309}
\la x^2 \ra \, = \frac{2T}{M} t^2 E_{2-\alpha,3}(-\gamma_\alpha
t^{2-\alpha})
 \APPROX{t\rightarrow \infty}\frac{2T}{\gamma_\alpha M}
\frac{t^\alpha}{\Gamma(1+\alpha)}\ . \ee In the last equation,
thermal initial conditions have been assumed ($x_0 =0$, $v_0^2 =
T/M$). Equation (\ref{eq309}) shows that the coupling to a fractal
heat bath leads in general to anomalous diffusion, $\la x^2 \ra
\sim t^\alpha$, $\alpha\neq 1$. The band form factor (\ref{eq181})
gives rise to subdiffusion when $\alpha\!<\!1$ and to
superdiffusion when $1\!<\!\alpha\!<\!2$. In general a wide range
of anomalous transport behaviors can be realized through RMT
(quantum chaotic) environments, with the character of the
diffusion related to the microscopic properties of the quantum
environment[20].

\section{Random Matrix Theory master equation:\\ weak coupling plus two baths}

We are now in a position to formulate the master equation for the quantum analog
of Fig. 1, a system coupled to two heat baths $B_1$ and $B_2$.
The corresponding Hamiltonian is given by
\be
H = H_S + H_{B_1}  + H_{B_2}+ Q_1(x)\otimes V_1 + Q_2(x)\otimes V_2 \ ,
\ee
where $H_S = \sum_i p_i^2/2m_i + U(x_i,x_j)$ is a N-particle system Hamiltonian,
$Q_1(x_1,..., x_N)$ and $Q_2(x_1,...,x_N)$ are two (arbitrary) system operators
and $V_1$ and $V_2$ are two random matrix bath operators. The variance of these
two random operators is given by Eq.~(\ref{eq3}) with the respective parameters
of the  two baths given by $T_k$, $\Gamma_k$ and ${\kappa_0}_k$, $k=1,2$. A quantum
master equation for the system can now be derived  using the method described in
section \ref{sec2}. This leads to the ensemble averaged equation
\bea
\frac{d\overline \rho_S(t)}{dt} &=& \!-i \,[H_S,\overline \rho_s(t)] \\
\!&-&\!\! \int_0^\infty d\tau \, [Q_1,[\widetilde{Q_1}(-\tau),\overline \rho_S(t)]]\,
\overline K'_1(\tau) -i \int_0^\infty d\tau  \,[Q_1,\{\widetilde{Q_1}(-\tau),\overline
\rho_S(t)\}] \, \overline K_1''(\tau)\nonumber \\
\!&-&\!\! \int_0^\infty d\tau \, [Q_2,[\widetilde{Q_2}(-\tau),\overline \rho_S(t)]]\,
\overline K'_2(\tau) -i \int_0^\infty d\tau  \,[Q_2,\{\widetilde{Q_2}(-\tau),\overline
\rho_S(t)\}] \, \overline K_2''(\tau ) \nonumber
\eea
In the following we take for the system  a 1D harmonic crystal of unit masses and
frequency $\omega$. We attach the two heat baths at both ends, $x_1$ and $x_N$,
of the linear chain. We thus have $H_S = \sum_i p_i^2 + \omega^2/2 \sum_{i,j}
G_{ij} x_i x_j$, where the matrix ${\bf G}$ is defined as  $G_{ij}= 2\delta_{ij}
-\delta_{i+1,j} -\delta_{i-1,j}$. The system operators are respectively given by
$Q_1(x_1,..., x_N) = x_1$ and   $Q_2(x_1,..., x_N) = x_N$. In the Ohmic regime,
the corresponding master equation can be easily written in coordinate representation in the form
\bea
\frac{d}{dt}\overline \rho_S(x_i,x_i',t)& =&\left\{ \frac{i}{2}\sum_i\Big(
\frac{\partial^2}{\partial x_i^2} - \frac{\partial^2}{\partial x_i'^2} \Big)
 -\frac{i}{2}\, \omega^2 \sum_{ij}G_{ij} \Big(x_ix_j - x'_ix'_j \Big)\right. \nonumber \\
& -& D_1\,(x_1-x_1')^2 -
\gamma_1\,(x_1-x_1')\Big(\frac{\pd}{\pd x_1}- \frac{\pd}{\pd x'_1}\Big)\nonumber \\
 & -& \left.D_2\,(x_N-x'_N)^2 -
\gamma_2\,(x_N-x_N')\Big(\frac{\pd}{\pd x_N}- \frac{\pd}{\pd x'_N}\Big)\right\} \overline \rho_S(x,x',t) \ ,
\eea
with $D_k =2T_k \gamma_k$. Taking the Wigner transform (\ref{eq1202}) we further obtain
\bea
\label{eq40}
\frac{\partial f}{\partial t}(q_i,p_i,t)&=& -\sum_{i=1}^N p_i\frac{\partial f}{\partial
  q_i}+ \sum_{i,j=1}^N \frac{\pd U}{\pd q_i} \frac{\partial f}{\partial p_j}
+ 2\gamma_1
 \frac{\partial }{\partial p_1}(p_1f) + 2\gamma_1 T_1 \frac{\partial^2
  f}{\partial p_1^2}\nonumber \\
&+& 2\gamma_2
 \frac{\partial }{\partial p_N}(p_Nf) + 2\gamma_2 T_2 \frac{\partial^2
  f}{\partial p_N^2} \ .
\eea
Introducing the  notation, $x_i= q_i$, $x_{i+N}=p_i$, $i=1,..., N$, Eq.~(\ref{eq40})
can be rewritten in the compact form
\be
\label{eq41}
\frac{\partial f}{\partial t}(x_i,t) = \sum_{i=1}^{2N} \frac{\pd}{\pd x_i}(\xi_i f)
 +\frac{1}{2} \sum_{i,j=1}^{2N} \frac{\pd^2}{\pd x_i \pd x_j}(d_{ij}f)\ ,
\ee
with $\xi_i =\sum_{ij} a_{ij}x_j$.
 The two  $2N\times 2N$ matrices {\bf a} and {\bf d}  are given by
\be
{\bf a}=
\left(\begin{array}{cr}
0 & -{\bf I} \\
\omega^2 {\bf G} & {\bf R}
\end{array}\right)
\hspace{.5cm}\mbox{ and } \hspace{.5cm} {\bf d} =
\left(\begin{array}{cc}
0 & 0 \\
0 & {\boldmath \varepsilon}
\end{array}\right) \ ,
\ee
where $0$ and $I$ denote the null and unit $N\times N$ matrices, and the two $N\times N$
matrices ${\bf R}$ and ${\boldmath \varepsilon}$ obey $R_{ij} = (2\gamma_1 \delta_{1i}+
2\gamma_2 \delta_{Ni})\delta_{ij}$ and $\varepsilon_{ij}=2 T_i R_{ij}$.
The classical generalized Klein--Kramers equation (\ref{eq41}) has been studied by
Rieder, Lebowitz and Lieb \cite{rie67} (the quantum problem has been treated  using a
Langevin approach  in \cite{zur90}). Of interest here is the stationary non--equilibrium
solution $\pd f_s(x_i)/\pd t = 0 $ for a small temperature difference,
 $T_1= T+\Delta T/2$, $T_2=T-\Delta T/2$, $\Delta T\ll T$. It is given by
\be f_s({\bf x})= (2\pi)^{-N} \mbox{Det}[{\bf
b}^{-\frac{1}{2}}]\exp\left[-\frac{1}{2} \sum_{i,j=1}^{2N}
b^{-1}_{ij} x_i x_j\right]\ , \ee where ${\bf b}$ is the $2N\times
2N$ covariance matrix, $b_{ij} = \la x_i x_j \ra = \int f_s({\bf
x}) x_i x_j d{\bf x}$. It is useful to write the covariance matrix
 in the form
\be {\bf b}= \left(\begin{array}{cc}
{\bf x} & {\bf z} \\
{\bf z}^\dagger & {\bf y}
\end{array}\right) \ ,
\ee
where the $N\times N$ matrices ${\bf x}$, ${\bf y}$ and ${\bf z}$ give,
respectively,  the correlations between the coordinates, the momenta and
between the coordinates and momenta, $x_{ij} = \la q_iq_j\ra$, $y_{ij} =
\la p_ip_j\ra$, $z_{ij} = \la q_ip_j\ra$. In the limit of large $N$ and
small coupling $\gamma=\gamma_1=\gamma_2$,  one finds
\be
{\bf x} = \frac{T}{\omega^2}\, {\bf G}^{-1},
 {\bf y} = \left(\begin{array}{cccc}
T(1)& & & \\
 &T(2)& & \\
& &\ddots &\\
& & &T(N)
\end{array}\right),
 {\bf z} = \left(\begin{array}{cccc}
0&\frac{\gamma \Delta T}{\omega^2} & & \\
 -\frac{\gamma \Delta T}{\omega^2}&0 &\frac{\gamma \Delta T}{\omega^2}& \\
&-\frac{\gamma \Delta T}{\omega^2} &\ddots &\frac{\gamma \Delta T}{\omega^2}\\
& &-\frac{\gamma \Delta T}{\omega^2} &0
\end{array}\right) \ .
\ee
Here $T(1) = T+\Delta T/2$, $ T(N) = T-\Delta T/2$ and $T(i) = T$ otherwise.
We see that the temperature profile in constant in the bulk and presents  a discontinuity at the edges.
By inverting the covariance matrix ${\bf b}$ we eventually arrive at
\be
\label{eq45}
f_s(q_i,p_i) \sim \exp \left[- \sum_i\frac{p_i^2}{2T(i)} -\frac{U}{2T}
 -\frac{\gamma \Delta T}{2T} \sum_i\frac{p_i}{\omega^2 T(i)}\left(\frac{\pd U}{\pd q_{i-1}}
 - \frac{\pd U}{\pd q_{i+1}}\right)\right]\ .
\ee This equation is valid in the limit $\Delta T, \gamma
\rightarrow 0$. We note that Eq.~(\ref{eq45}) reduces to the
equilibrium Boltzmann distribution when $\Delta T =0$. The last
term in (\ref{eq45}) is proportional to the heat flux $J = \gamma
\Delta T/2$ (note that $J$ is not proportional to the temperature
gradient). Taking the inverse Wigner transform we obtain the
stationary density operator in the form \be \overline
\rho_S(q_i+\frac{r_i}{2}, q_i-\frac{r_i}{2}) = \prod_i \exp\left
[\frac{1}{2 T(i)} \left (\frac{\gamma \Delta T}{2
T}\frac{1}{\omega^2}\left(\frac{\pd U}{\pd q_{i-1}} - \frac{\pd
U}{\pd q_{i+1}}\right)-iT(i) r_i \right)^2\right] \exp\left[
-\frac{U}{2T}\right ]\ . \ee
This form of the solution gives us a
guide on how to better solve the general problem of Fig. 1 using a
general quantum master equation.

\section{Conclusions}

The transport properties of classical and quantum systems in
non-equilibrium steady states was formulated and discussed. While
the situation for classical non-equilibrium systems does not seem
very well understood yet, it is clearly far better understood than
the quantum counterpart. At the moment one can formulate a master
equation for the non-equilibrium density matrix, however, aside
from the trivial harmonic case discussed above, very little is
known. It would be very desirable to see if one can derive
Fourier's law or the temperature profile from Random Matrix
models, and generally how far one can push the matrix models which
seem to provide a tractable approach to quantum thermalization.

This work was supported by the Office of Naval Research under
contract \# N00014-01-1-0594. DK thanks R. Olkiewicz and P.
Garbaczewki, the organizers of the 38$^{th}$ Winter School of
Theoretical Physics.

\end{document}